\documentclass[aps,prl,twocolumn]{revtex4}
\usepackage{graphicx}
\usepackage{epsfig}

\begin{document}

\title{Measurement of the Current-Phase Relation of SFS $\pi$-Josephson junctions}

\author{S. M. Frolov}
\author{D. J. Van Harlingen}

\affiliation{Department of Physics, University of Illinois at
Urbana-Champaign, Urbana, IL, 61801, USA}

\author{V. A. Oboznov}
\author{V. V. Bolginov}
\author{V. V. Ryazanov}

\affiliation{Institute of Solid State Physics, Russian Academy of
Sciences, Chernogolovka, 142432, Russia}

\date{\today}

\begin{abstract}

\noindent We present measurements of the current-phase relation
(CPR) of Superconductor-Ferromagnet-Superconductor (SFS) Josephson
junctions as a function of temperature. The CPR is determined by
incorporating the junction into a superconducting loop coupled to
a dc SQUID, allowing measurement of the junction phase difference.
Junctions fabricated with a thin ($\sim 22$ nm) barrier of
Cu$_{0.47}$Ni$_{0.53}$ sandwiched between Nb electrodes exhibit a
re-entrant critical current with temperature, vanishing at $T
=T_\pi \sim 2-4$ K. We find that the critical current is negative
for $T < T_\pi$, indicating that the junction is a $\pi$-Josephson
junction.  We find no evidence for second-order Josephson
tunneling near $T_\pi$ in the CPR predicted by several theories.
\end{abstract}


\maketitle The interplay between superconductivity and magnetism
in thin film Superconductor-Ferromagnetic (SF) structures has long
attracted substantial theoretical and experimental attention. Over
twenty years ago, it was predicted that a
Superconductor-Ferromagnet-Superconductor (SFS) Josephson junction
could become a $\pi$-junction, characterized by a minimum
Josephson coupling energy at a phase difference of $\pi$, due to
exchange field-induced oscillations of the order parameter (OP) in
the ferromagnetic barrier \cite{Buzdin}.  Such $\pi$-junctions
were only achieved recently in systems with weak ferromagnetic
barriers and demonstrated by conventional transport
\cite{Ryazanov-LT, Ryazanov-junction, Aprili-junction, Sellier,
Blum} and SQUID interference \cite{Ryazanov-triangle,
Aprili-squid, Aprili-Hall} measurements. $\pi$-Josephson behavior
has also been reported in mesoscopic Superconductor-Normal
metal-Superconductor (SNS) junctions driven into a nonequilibrium
state by the injection of quasiparticles into the barrier
\cite{Klapwijk-junction, Birge-junction}, in nanoscale cuprate
grain boundary junctions for which the supercurrent is dominated
by zero-energy Andreev bound states induced by the d-wave order
parameter \cite{Testa}, and in nanoscale constrictions in
superfluid $^3$He \cite{Packard}. Because they result in a
doubly-degenerate phase potential when incorporated into a
superconducting loop, $\pi$-junctions have been proposed as
building blocks for superconducting qubits \cite{Ioffe}.

In this paper, we present measurements of the current-phase
relation (CPR) of an SFS Josephson junction that demonstrate
directly the sign change in the critical current when the junction
undergoes a crossover into a $\pi$-state below a temperature
$T_\pi$ at which the critical current vanishes. We investigate the
crossover region near $T_\pi$ for which it has been predicted that
second-order Josephson effects should dominate, yielding a sin($2
\phi$) CPR \cite{Chtchelkachev, Radovic}. We find no evidence for
such a contribution.

\begin{figure}[b]
 \centering
 \includegraphics[width=5cm,bb=120 60 520 710]{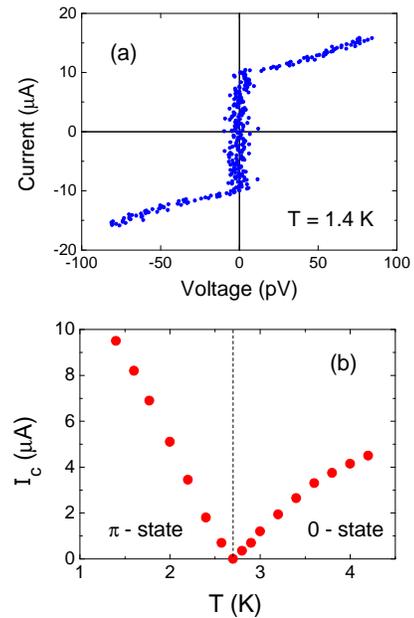}
 \caption{(a) Current vs. voltage for a Nb-CuNi-Nb Josephson junction
 measured at $T=1.4$ K. (b) Variation of the critical current with
 temperature showing re-entrance at $T \approx 2.7$ K characteristic of a transition into
 a $\pi$-junction state. \label{SFS-IV-IcT}}
\end{figure}

A $\pi$-junction is a Josephson junction with a negative critical
current $I_c$.  Thus, the current $I_J$ through a $\pi$-junction
for a given superconducting phase difference across the junction
$\phi$, assuming a purely sinusoidal form for the CPR, is given by
$I_J(\phi) = - |I_c| ~ $sin$\phi = |I_c| ~ $sin$(\phi + \pi)$, in
terms of the magnitude of the critical current $|I_c|$. The
minimum energy state of an isolated $\pi$-junction corresponds to
a phase shift of $\pi$ across the junction \cite{Bulaevskii}, in
contrast to an ordinary Josephson junction, or 0-junction, for
which the minimum energy is at zero phase difference.

In SF bilayer structures, superconducting correlations are known
to exist in the F-layer due to the proximity effect.  Because of
the exchange field energy $E_{ex}$, Cooper pairs in the F-layer
have non-zero center-of-mass momentum $Q = 2E_{ex}/{\hbar}v_F$,
where $v_F$ is the Fermi velocity.  The wave function of these
Cooper pairs at distance $x$ from the SF interface obtains a phase
multiplier exp$(+iQx)$ or exp$(-iQx)$, depending on the
orientation of the electron spins. Taking into account all spin
states, the OP $\Psi$ induced in the F-layer, has the form:

\begin{eqnarray}
\Psi (x) &\sim&  \textrm{cos} \left( \frac{x}{\xi_{F2}} \right) \:
\textrm {exp} \left( -\frac{x}{\xi_{F1}}\right), \label{eq:psi}
\end{eqnarray}
which describes the decay of the OP in the ferromagnetic layer
over length $\xi_{F1}$, modulated by spatial oscillations with the
period $2 \pi \xi_{F2}$. In the dirty limit, $\xi_{F1}$ and
$\xi_{F2}$ are given by \cite{Ryazanov-junction}:

\begin{eqnarray}
\xi_{F1,F2} = \left\{ \frac{\hbar D}{ \left[ (\pi k_B T)^2 +
E_{ex}^2 \right]^{1/2} \pm \pi k_B T}\right\}^{1/2}
,\label{eq:ksi}
\end{eqnarray}
where $D$ is the diffusion constant. Such oscillations of the OP
have been confirmed in SF-bilayers by measurements of the
superconducting critical temperature \cite{Jiang} and by tunneling
spectroscopy \cite{Aprili-tunneling}.

In SFS junctions, the OP oscillations cause the magnitude of the
critical current to vary with the barrier thickness, vanishing at
one or more thicknesses \cite{Aprili-junction, Sellier, Blum}.  A
ferromagnetic layer with thickness of order 1/2 (or other odd
half-integer value) of the oscillation wavelength results in a
sign change in the OP between the superconductor electrodes,
meaning that the junction becomes a $\pi$-junction. Although this
condition can be achieved with ultrathin barriers of a strong
ferromagnet \cite{Blum}, it is experimentally advantageous to use
thicker barriers of a weakly-ferromagnetic alloy. Ferromagnetic
layers with thicknesses in the range 10-30 nm are ideal because
they are thick enough to form a uniform Josephson barrier yet thin
enough to allow a measurable supercurrent.  For $\xi_{F1}$ and
$\xi_{F2}$ to be in the appropriate range, the Curie temperature
of the ferromagnetic material should be of order $20-100$ K. SFS
$\pi$-junctions of this type have been fabricated using metallic
alloys consisting of the strong ferromagnet Ni diluted with either
diamagnetic Cu \cite{Ryazanov-junction} or paramagnetic Pd
\cite{Aprili-junction}.

As can be seen from Eq.(\ref{eq:ksi}), both $\xi_{F1}$ and
$\xi_{F2}$, and hence the junction critical current, vary with
temperature. Another advantage of a weak-ferromagnetic barrier is
that $E_{ex}$ can be made comparable to $k_B T$ in the
experimentally-accessible temperature range ($1-4$ K) so that the
changes in $\xi_{F1}$ and $\xi_{F2}$ are maximized. This allows an
SFS Josephson junction of appropriate barrier thickness to be
tuned between the $0$ and $\pi$ states via temperature, enabling
the crossover region to be explored in a single junction. We
utilize this capability in our experiments.

Our SFS junctions were prepared in a multi-step process by optical
lithography and magnetron sputtering.  The base and top
superconducting layers are dc-sputtered Nb with thicknesses $100$
nm and $240$ nm respectively, separated by a $22$ nm barrier layer
of rf-sputtered {Cu$_{0.47}$Ni$_{0.53}$}, a weakly-ferromagnetic
alloy which has a Curie temperature of $\sim 60$ K. The size of
the junctions was $50 ~\mu$m$~\times~ 50 ~\mu$m, as defined by a
window in an insulating SiO layer deposited directly on top of the
CuNi.

Because of the low normal state resistance of the SFS junctions
($R_N \sim 10 ~\mu \Omega$), the current-voltage characteristics
of the SFS junctions are measured using a SQUID potentiometer
setup. In Fig. \ref{SFS-IV-IcT}(a), we show a typical $I$ vs. $V$
curve from which the critical current is determined. From a series
of these curves, the critical current is plotted as a function of
temperature, as in Fig. \ref{SFS-IV-IcT}(b). As the temperature is
lowered from $4.2$ K, the critical current decreases, vanishes at
a temperature {$T_ \pi = 2.75$ K}, and then increases again. This
re-entrance is consistent with a transition between $0$-junction
and $\pi$-junction states \cite{Ryazanov-junction, Sellier}. At
the maximum critical current $\sim 10 ~\mu$A, the product $I_c R_N
\approx 100 $ pV.

\begin{figure}[b]
 \centering
 \includegraphics[width=6cm,bb=130 100 550 720]{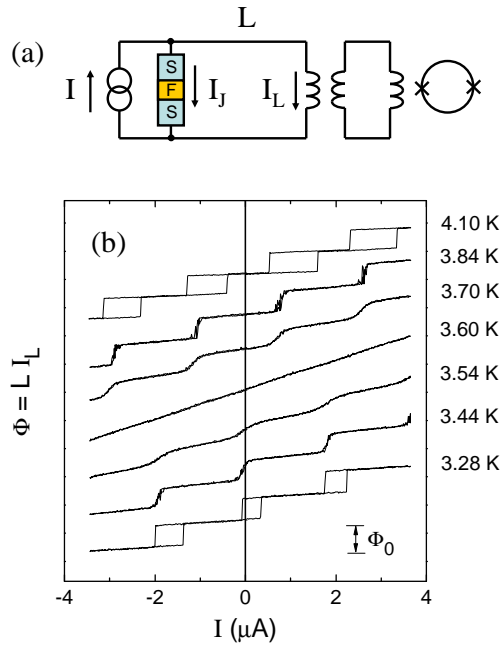}
 \caption{(a) Circuit for measuring the current-phase relations of an SFS junction.
(b) Magnetic flux $\Phi$ in the rf SQUID loop vs. applied current
$I$ showing a transition from hysteretic to non-hysteretic curves
as $|I_c|$ drops. Curves offset for clarity.\label{SFS-PhiI-all}}
\end{figure}

\begin{figure}[b]
 \centering
 \includegraphics[width=5.5cm,bb=130 130 480 720]{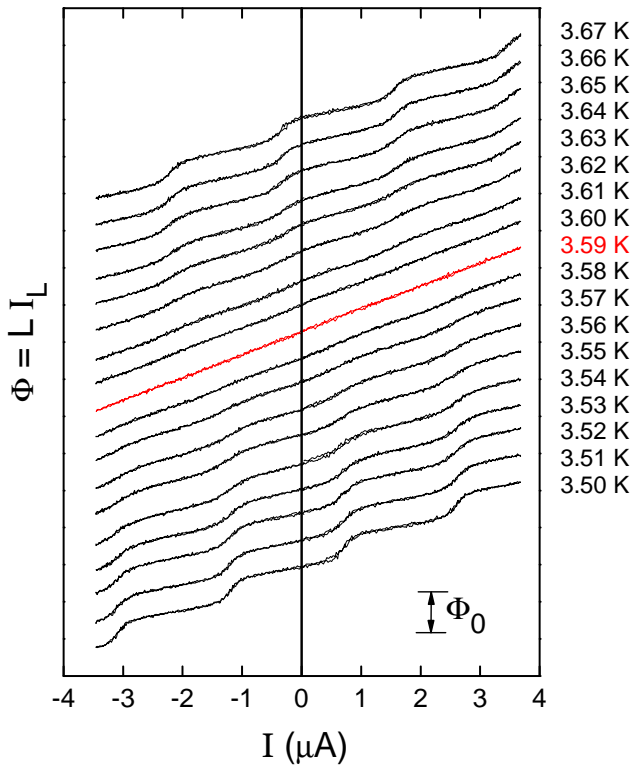}
 \caption{Modulation of the magnetic flux in the rf SQUID loop as a
 function of current applied across the SFS junction for a series of
 temperatures. As the temperature is lowered, the critical current vanishes
 at $T=3.59$ K, below which the modulation shifts phase by $\pi$.
 Curves offset for clarity. \label{SFS-PhiI}}
\end{figure}

Conventional measurements of the current-voltage characteristic of
the junction are not sensitive to the sign of the critical current
nor to the shape of the current-phase relation; only $|I_c|$ can
be determined. To verify $\pi$-junction behavior, it is necessary
to perform a phase sensitive measurement by including the junction
in a multiply-connected geometry.  The sign of the critical
current can be detected in an rf SQUID configuration by shorting
the electrodes of the junction with a superconducting loop. For
sufficiently high inductance L (such that $|\beta_L| = 2 \pi |I_c|
L / \Phi_0 \gg 1)$, a loop containing a $\pi$-junction in zero
applied magnetic field will exhibit a spontaneous circulating
current, generating a magnetic flux of $(1/2) \Phi_0$ in the loop
which can be detected by a SQUID magnetometer or Hall probe; for
smaller inductance, such that $|\beta_L| < 1$, it is energetically
favorable to flip the phase of the junction into its high energy
state $\phi = 0$ in which there is no circulating current.
Alternatively, the junction can be connected in parallel with a
conventional Josephson junction to form a dc SQUID. In this case,
a $\pi$-junction is identified by a minimum of the SQUID critical
current in zero magnetic field. We note that measurements of both
the minimum in the critical current in dc corner SQUIDs
\cite{Wollman} and the spontaneous flux in tricrystal rings
\cite{Kirtley} have been used to demonstrate a similar but
distinct effect, the phase shift of $\pi$ between orthogonal
directions in the d-wave superconducting cuprates.

The most complete way to characterize the $\pi$-junction behavior
is to measure the current-phase relation (CPR).  The CPR specifies
the magnitude and sign of the sinusoidal component of the critical
current as well as the amplitudes of any higher harmonics that may
be present.  The CPR can be measured in the rf SQUID configuration
shown in Fig. \ref{SFS-PhiI-all}(a). A dc SQUID galvanometer is
used to measure the current $I_L$ that flows through the
superconducting loop as a function of the current $I$ applied
across the junction. The CPR function $I_J(\phi)$ is related to
$I$ and $I_L$ by

\begin{eqnarray}
I &=& I_J (\phi) +I_L = I_J \left(\frac{2 \pi \Phi}{\Phi_0}\right)
+ \frac{\Phi}{L},\label{eq:rfSQUID}
\end{eqnarray}
where $\Phi$, the total magnetic flux in the loop, is related to
the junction phase $\phi = 2 \pi \Phi/\Phi_0$ by the phase
constraint around the rf-SQUID loop, and to $I_L = \Phi/L$
provided that there is no external flux linking the SQUID loop.

\begin{figure}[b]
 \centering
 \includegraphics[width=5.5cm,bb=130 255 670 720]{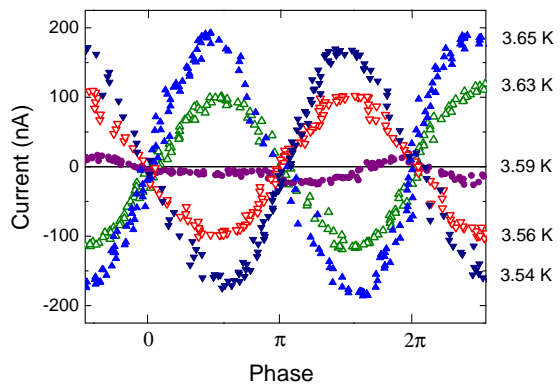}
 \caption{Current-phase relation derived from the rf SQUID modulation
 curves of Fig. \ref{SFS-PhiI} showing the transition to a $\pi$-Josephson
 junction as the temperature is lowered.
 \label{SFS-CPR}}
\end{figure}

\begin{figure}[b]
 \centering
 \includegraphics[width=5.5cm,bb=150 200 680 720]{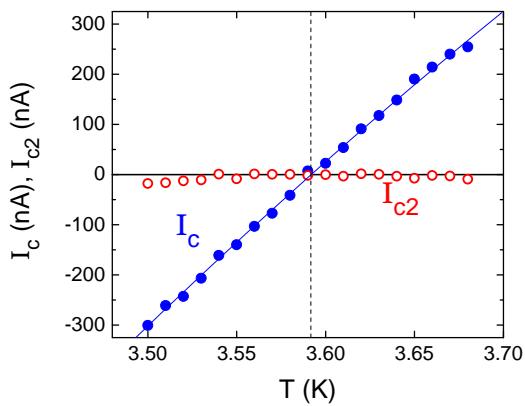}
 \caption{Variation of $I_c$ and $I_{c2}$, the sin$\phi$ and sin($2 \phi$)
 components of the Josephson critical current, with temperature,
 showing the sign change in $I_c$ and absence of a significant
 $I_{c2}$. \label{SFS-IcT-Ic2T}}
\end{figure}

For our phase-sensitive measurements, the SFS $\pi$-junction is
incorporated into an rf-SQUID loop with inductance $L \approx 1$
nH.  This loop is fabricated in the shape of a planar washer to
which a coil of Nb superconducting wire is coupled and connected
to the input terminals of a commercial dc SQUID sensor. As current
$I$ is applied across the SFS junction, the magnetic flux in the
loop is modulated due to the winding of the phase of the Josephson
junction according to Eq.(\ref{eq:rfSQUID}).  The inductance $L$
determines the critical current range (here up to $\sim 300$ nA)
over which the rf-SQUID response remains non-hysteretic
($|\beta_L| < 1$) so that the full CPR period can be mapped out.

For one sample, a series of curves plotting the flux in the
rf-SQUID loop $\Phi$ vs. applied current $I$ for different
temperatures is shown in Fig. \ref{SFS-PhiI-all}(b). We note that
the flux axis is self-calibrating since each period corresponds to
a one flux quantum $\Phi_0$ change in the loop flux. Plotted in
this form, the overall slope of the curves is $L$, the loop
inductance, which is determined to be $1.28 \pm 0.01$ nH.  The
curves are strongly hysteretic at $T = 4.2$ K and at low
temperatures. They become non-hysteretic in the temperature range
from $3.7-3.5$ K. At $T \approx 3.6$ K, there is no discernible
modulation in $\Phi$ indicating that $I_c = 0$, and we identify
this as the $0-\pi$-junction crossover temperature $T_\pi$.  All
of the SFS junctions that we have studied were fabricated to
exhibit a crossover temperature between $0$ and $\pi$-states in
the range $2-4$ K.

Figure \ref{SFS-PhiI} shows in detail the temperature range for
which $-1 \leq \beta_L \leq 1$. The modulation of the flux is now
more accurately seen to disappear at T$_\pi = 3.59$ K. The most
striking feature of the data in Figs. \ref{SFS-PhiI-all} and
\ref{SFS-PhiI} is that the relative phase of the modulation
abruptly changes by $\pi$ as the temperature is varied from above
to below $T_\pi$. Due to the presence of stray residual magnetic
fields ($\sim 10$ mG) in the cryostat, the phase of the modulation
(and hence the junction phase difference) is not in general zero
for zero applied current and varies slightly with temperature.
This background phase shift is roughly linear in the vicinity of
the $0-\pi$ transition.

As can be seen in Eq.(\ref{eq:rfSQUID}), the current-phase
relation can be directly extracted from the data in Fig.
\ref{SFS-PhiI} by subtracting the linear flux term and taking
account of any phase shifts arising from background fields. The
CPR for several temperatures near $T_\pi$ is shown in Fig.
\ref{SFS-CPR}.  The CPR has a sinusoidal form.  No doubling of the
periodicity is observed in the CPR at any temperature, suggesting
that second-order Josephson tunneling harmonics, if present, never
dominate the CPR of the junction.  At $T_\pi = 3.59$ K, only
aperiodic fluctuations of the current are observed, which limit
the resolution of our critical current measurements to $\sim 10$
nA. The CPR curves for temperatures above and below $T_\pi = 3.59$
K are out of phase by $\pi$, verifying that the critical current
of the SFS Josephson junction changes sign at $T_\pi$.

The critical current as a function of temperature can be extracted
from the CPR curves, or, more accurately, directly from the family
of curves in Fig. \ref{SFS-PhiI} by fitting them to the form of
Eq.(\ref{eq:rfSQUID}). For the CPR, we assume the functional form
{$I_J(\phi)=I_c$ sin($\phi$) + $I_{c2}$ sin($2 \phi$)}, allowing
for a second-order Josephson component. $I_{c}$ and $I_{c2}$
determined from the fits are plotted in Fig. \ref{SFS-IcT-Ic2T}.
The temperature variation and sign change of $I_{c}$ are clearly
seen. $I_{c2}$ is relatively flat, never exceeding a few percent
of the maximum sinusoidal component $I_c$ and, more significantly,
vanishes along with $I_c$ at $T_\pi$. This suggests that the
induced $I_{c2}$ value is likely an artifact of the fitting
procedure rather than a physical second-order Josephson component
in the CPR.

The sin($2 \phi$) component has been predicted to persist and
dominate the CPR at the crossover point, inhibiting the critical
current from vanishing completely at $T_\pi$ \cite
{Chtchelkachev}. In contrast, our results indicate that the
critical current is zero at $T_\pi$ and that a sin($2 \phi$) term
is not present in the CPR for these SFS junctions.  It should be
emphasized that the data presented in this paper is obtained on
SFS Josephson junctions in the dirty limit $\ell < \xi_{FR}$,
where $\ell$ is the mean free path in the ferromagnetic barrier.
Some models predict that the barrier must be in the clean limit
for second-order Josephson tunneling to dominate the CPR near the
$0-\pi$ transition \cite{Radovic}.  In support of this, we note
that a sin($2 \phi)$ component has been observed in
voltage-controlled SNS Josephson junctions in the ballistic regime
\cite{Klapwijk-squid}. Thus, it is possible that a sin($2 \phi)$
term may arise in samples in the clean limit or with higher
interface transparency.

In conclusion, we have performed phase-sensitive measurements on
SFS Josephson junctions that exhibit a transition from a $0$-state
into a $\pi$-state at a crossover temperature $T_\pi$. The
current-phase relation of the junctions is mapped out as a
function of temperature, demonstrating the vanishing of $I_c$ at
$T_\pi$ and the sign change in the critical current at this
temperature.  No higher-order harmonics in the CPR are observed
for these junctions.

We thank Marco Aprili and Alexander Golubov for useful
discussions. Work supported by the National Science Foundation
grant EIA-01-21568 and by the U.S. Civilian Research and
Development Foundation (CRDF) grant RP1-2413-CG-02. We also
acknowledge extensive use of the Microfabrication Facility of the
Frederick Seitz Materials Research Laboratory at the University of
Illinois at Urbana-Champaign.

\bibliography{sfs}

\end{document}